# High values of Ferroelectric polarization and magnetic susceptibility in BFO nanorods

**Nabanita Dutta[1], S.K.Bandyopadhyay[1*], Subhasis Rana[1], Pintu Sen[1], A.K.Himanshu[1] and P.K.Chakraborty[2]**

1. **Variable Energy Cyclotron Centre, 1/AF, Bidhan Nagar, Kolkata-700 064, India.**
2. **Department of Physics, Burdwan University, Burdwan 713104, India.**

***Corresponding author. E-mail: skband@vecc.gov.in**

**Abstract:**

Remarkably high values of polarization as well as a significant magnetic susceptibility have been observed in multiferroic Bismuth Ferrite (BFO) in the form of nanorods protruding out. These were developed on porous Anodised Alumina (AAO) templates using wet chemical technique. Diameters of nanorods are in the range of 20-100 nm. The high values of polarization and magnetic susceptibility are attributed to the BFO nanorod structures giving rise to the directionality. There is no leakage current in P-E loop examined at various frequencies. Magnetocapacitance measurements reflect a significant enhancement in magnetoelectric coupling also.

**Keywords: A.** Bismuth Ferrite; A. Nanorod; E.Polarization; E. Magnetic susceptibility

## 1. Introduction:

Bismuth ferrite (BFO) is a quite interesting multiferroic showing coexistence of ferroelectric as well as antiferromagnetic properties [1,2]. BFO exists in nature as a partially covalent oxide with a rhombohedrally distorted perovskite structure belonging to the space group of *R3c* which allows for a ferroelectricity below 1083 K and G type antiferromagnetism with a very high Neel temperature $T_N$ of 643 K [3]. However, sometimes it exhibits weak ferromagnetism as a result of size manifestation that leads to a relative ordering. This type of weak ferromagnetism in antiferromagnets is explained through Dzialoshinsky-Moriya interaction, exchange bias, strain manipulation etc [4,5]. BFO seems to be a promising candidate for magnetic storage or other devices in spintronics applications [6]. The presence of $6s^2$ lone pair of electrons in bismuth accounts for the polarization for ferroelectricity. However, leakage currents due to oxygen vacancies or impurities of BFO lead to a major shortcoming. Recent approaches on it are being focused on single/polycrystals and substrate-free nanostructures as well as low-dimensional nanostructures, like nanoparticles, nanowires/rods fiber etc. In particular, the nanostructured forms of this material are expected to deliver enhanced properties of magnetization and polarization necessitating the drive for synthesizing in nanostructured form. However, each synthesis protocol has got some advantages and limitations. We have chosen Anodized Alumina (AAO) templates to grow BFO nanowires. In this communication, we report the development, characterization of BFO nanorods by template assisted wet chemical technique and their magnetic susceptibility and polarization studies.

## 2. Experimental details:

Commercially available AAO templates of 60µm thickness and 13mm dia with pore size distribution ranges from 20-200nm have been employed. 0.1M aqueous solutions of $Bi(NO_3)_3$ and $Fe(NO_3)_3$ were prepared with stoichiometric amounts of the nitrates using methoxymethanol as solvent. pH of the solution was adjusted to 2-3. Since the filling takes place with capillary action, it gets hindered due to very small diameter of the nanopores leading to inadequate filling of pores. The filling of pores was achieved by the directional flow of the ions in the template adopting the controlled vacuum technique. The templates with pores containing solution were dried in air and sintered for 3 hours at $750^0$C to get the required phase without unwanted grain growth. The templates are very much fragile and get bent during sintering. Hence we took special care keeping them sandwiched between two thin ceramic plates to receive in intact form.

The most important part of the process is the etching to observe nanowires/rods by electron microscopy. Templates were half etched so that the rods would be visible. They were not self supporting. We attempted controlled etching process with 1M NaOH and the bundle of nanowires emerged. Weight of BFO nanorod developed was measured using a very sensitive microbalance (with resolution of 1µgm) by subtracting the weight of AAO and it was in the range 80-100µgms. for various templates.

The nanowires and nanorods were examined by FEI Scanning Electron Microscope (SEM) with a resolution of 6nm and provided with Energy Dispersive X-ray (EDX) Analysis. Transmission electron microscopy (TEM) was taken by high resolution TEM (Model: FEI T20 with applied voltage of 200KV). Since the nanowires/ rods are grown in AAO templates, the sample preparation for TEM analysis is the crucial matter and it was subjected to extreme care through a number of steps. The AAO templates containing BFO nano wires half etched with NaOH were washed thoroughly to remove NaOH before TEM studies. Physical properties of synthesized BFO nano wires basically ferroelectric measurements were carried out to investigate ferroelectric behavior. Ferroelectric polarization was measured through P-E loop analyzer (Model: Precession, Radiant Technology) at different electric fields and frequency. Up to a field of 60KV/cm was employed without any break down. Magnetic susceptibility was measured from room temperature down to 25K at a field of 1300 gauss.

### 3. Results and Discussions:

#### 3.1. Electron Microscopy:

We have received two kinds of 1D nanostructure in the form of nanowires as well as nanorods as apparent from SEM similar to earlier groups [7, 8]. Cross sectional view is showing development of nanorods in Fig. 1a- a representative case. It demonstrates the structures of several nanorods (around 20 in a region of 5 µm x 5 µm) with high aspect ratio protruding out of the pores after partial etching. There were almost similar number of nanorods embedded in the pores. We noticed around 400 pores in 5 µm x 5 µm. TEM reveals nanorods of high density with intact structure in Fig. 1b. A clear SAED pattern (inset of Fig. 1b) with a prominent ring signifies development of polycrystalline BFO.

#### 3.2. Polarization:

Fig. 2 depicts P-E loops at various frequencies from 90-400 Hz. The hysteresis loops are quite appreciable characteristic of a ferroelectric material with the loop width decreasing with increasing applied frequencies. We have not attained saturation up to the applied voltage field of 60KV/cm. Measurement has been done for blank AAO template too and there is no signature of any hysteresis loop. The gap in the plot is seen due to improper depolarization.

The relative displacements of Bi and O ion induced by the repulsion of Bi(III) $6s^2$ lone pair of electrons result in a net polarization. Such polarization is an example of orientational polarization which is affected by distribution of the polarization domains. Relatively low value of polarization is a characteristic of nanoparticles. However, it is quite high in epitaxial BFO films and BFO single crystals [9], where the projection of polarization takes the orientation along the easy axis of growth of crystals. The net polarization matches along the direction of growth basically along (111) plane in case of BFO [10].

Most significantly, the observed polarization of BFO nanorods is remarkably high ($\sim 0.04 \mu C/cm^2$) considering the weight of the nanorods (80 micrograms). The observed polarization value comes in the same order as that of the single crystal. The value is substantially large as compared to our earlier polarization values obtained in case of agglomerated BFO nanoparticles [11]. This high value of polarization can be attributed to the directional growth of BFO as nanorod structure. We have employed a moderate sintering time to avoid the grain growth. Moreover, preheated furnace was employed to get rid of the loss of Bismuth. Bismuth to Iron ratio was nearly 1:1 as verified through EDX studies. There is no leakage current (indicated by decrease in polarization with the electric field) associated with our nanorods. The leakage current is observed mainly in grain boundary region due to the accumulation of defects as oxygen vacancies as well as metallic impurities like $Bi^{3+}$, iron oxide etc. Organized growths of BFO nanorods facilitate minimization of grain boundaries as well as absence of oxygen vacancies, metallic impurities leading to the disappearance of the leakage current.

### 3.3. Magnetic measurements:

Magnetic susceptibility versus temperature of BFO nanorods is shown in Fig. 3. BFO nanorods show a substantially high magnetic susceptibility of the order of 2.5emu/mole at low temperature under the application of magnetic field of 1300 Gauss. This result corresponds to the susceptibility obtained in single crystal reported earlier [12]. High susceptibility of BFO in such nanostructured form is attributed to the directional growth.

Usually a weak ferromagnetism is observed in bulk BFO, which gets enhanced in case of single crystal due to the matching of magnetization direction and easy axis of growth. The growth directionality of BFO nanorods in this case can be compared to the directionality in signal crystal leading to the high value of susceptibility. The negative susceptibility at room temperature is due to the diamagnetic contribution of AAO template, which is large compared to the weak ferromagnetic susceptibility of BFO nanorods at that temperature.

### 3.4. Magneto capacitance:

Magnetocapacitance gives an idea about the magnetoelectric coupling in an indirect way [13]. The capacitance per unit field has been plotted against the magnetic field as depicted in Fig. 4. Magnetocapacitance has been normalized with respect to the magnetic field to attain quantitative information and it appears to be quite high. Generally magnetocapacitance increases with the increasing value of the field and this signature is also evident in our measurement. In case of epitaxially grown thin films, if the polarization direction is in plane with the magnetic field, the film experiences maximum effect of magnetic field, but if it is out of plane, the net effect is zero. Similarly in this case also it becomes large due to the directional growth of the nano rod depending upon their orientation with respect to the magnetic field. The increasing capacitance with the increase of applied field suggests an indication of magnetoelectric coupling. Magnetocapacitance value is also high by a factor of five compared to that of BFO nanoparticles observed by us earlier [11] indicating a larger magnetoelectric coupling in case of nanorods. Thus, the nanostructure has not only enhanced the magnetic susceptibility and polarization, but it has also led to the enhancement of the magnetoelectric coupling. A smooth pattern is not noticed here due to the unequal interaction of nanorods. Each individual nanorod responds to the magnetic field differently as per their area and this leads to a random effect in magnetocapacitance.

## 4. Conclusion:

We have studied magnetic susceptibility as a function of temeperature, ferroelectric polarization and magnetocapacitance of BFO nanorods developed on AAO templates at various fields and frequencies. The values of polarization (~0.04μC/cm$^2$) are remarkably high compared to the bulk considering the low wt. of 80 μgms. Furthermore, the magnetic susceptibility is significantly high. This high value of polarization and magnetic susceptibility as well can be attributed to the directionality achieved in the nanorod structure. The pattern of the growth prohibits development of grain boundaries paving the pathway for leakage current free material. The magnetoelectric coupling in BFO nanorods as obtained through magnetocapacitance measurements is also quite high as compared to that of BFO nanoparticles. All measurements consistently establish the enhancement of multiferroic properties in the nanostructure, which is very important from the applications point of view.

**Acknowledgements:** Authors gratefully acknowledge Dr. P.K.Mukhopadhyay and Mr. Sakti Nath Das of S.N.Bose Center for basic Sciences for SEM studies and Mr. Pulak Kumar Roy of Saha Instititute of Nuclear Physics for TEM studies.

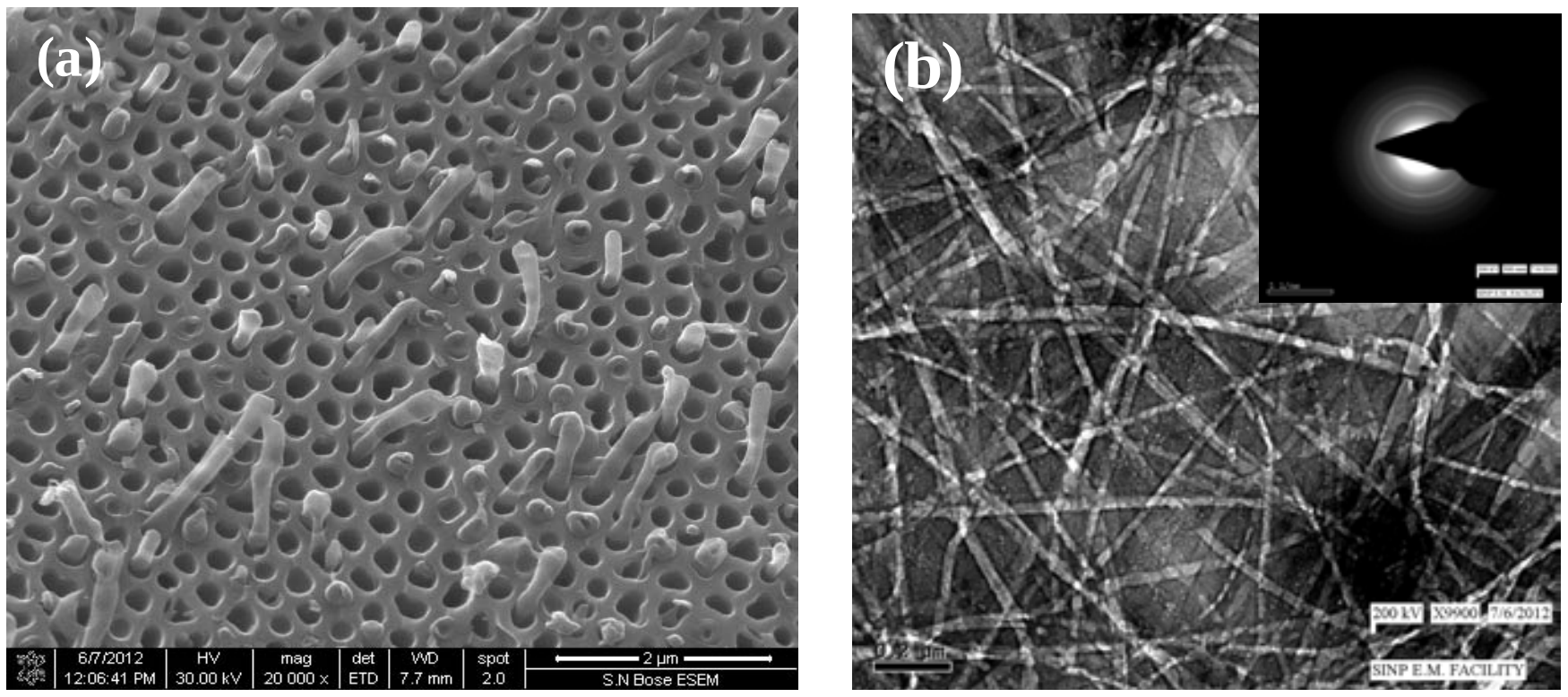

Fig. 1(a) SEM of nanorod protruding from pores and (b) TEM of BFO nanorods and corresponding SAED pattern shown in the inset.

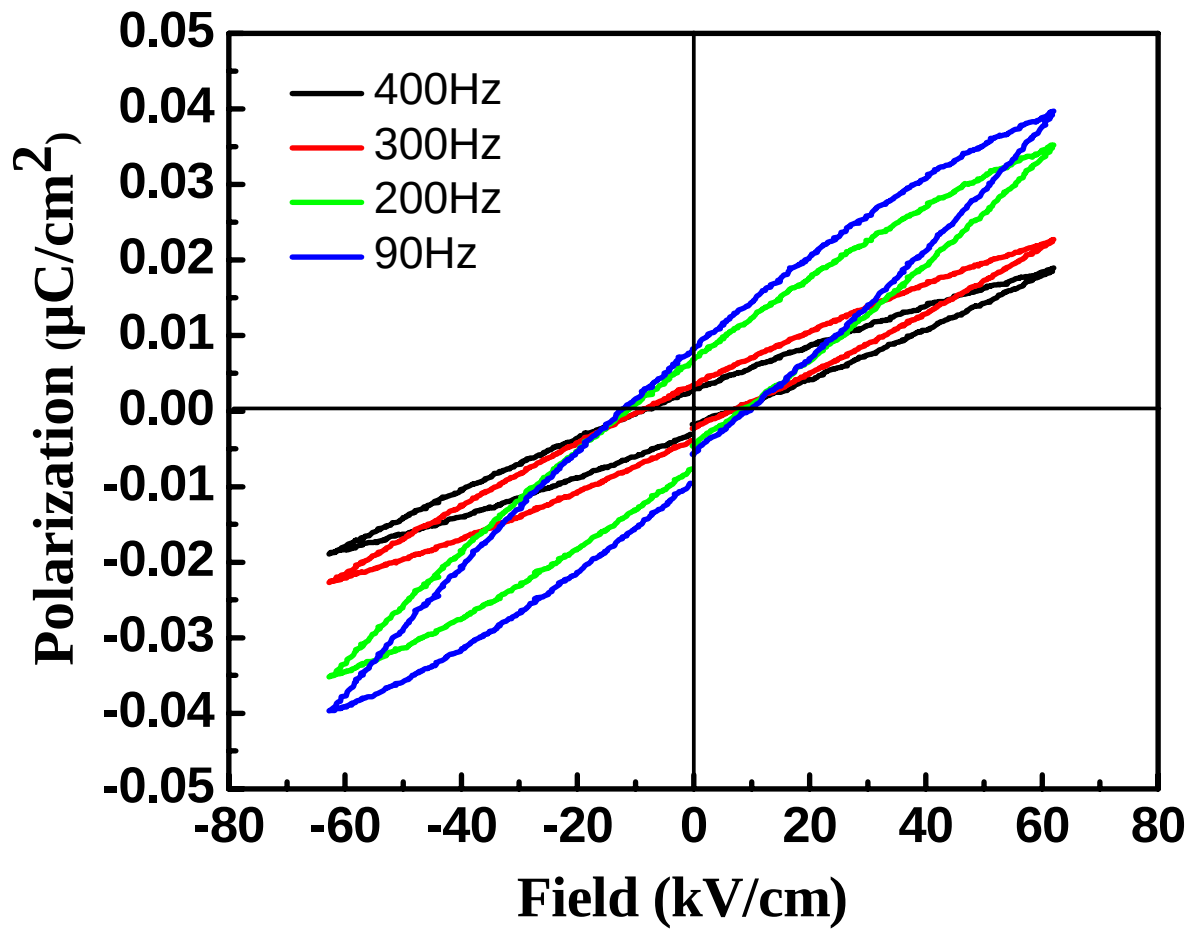

Fig. 2. Ferroelectric hysteresis loop of BFO nanorod.

\

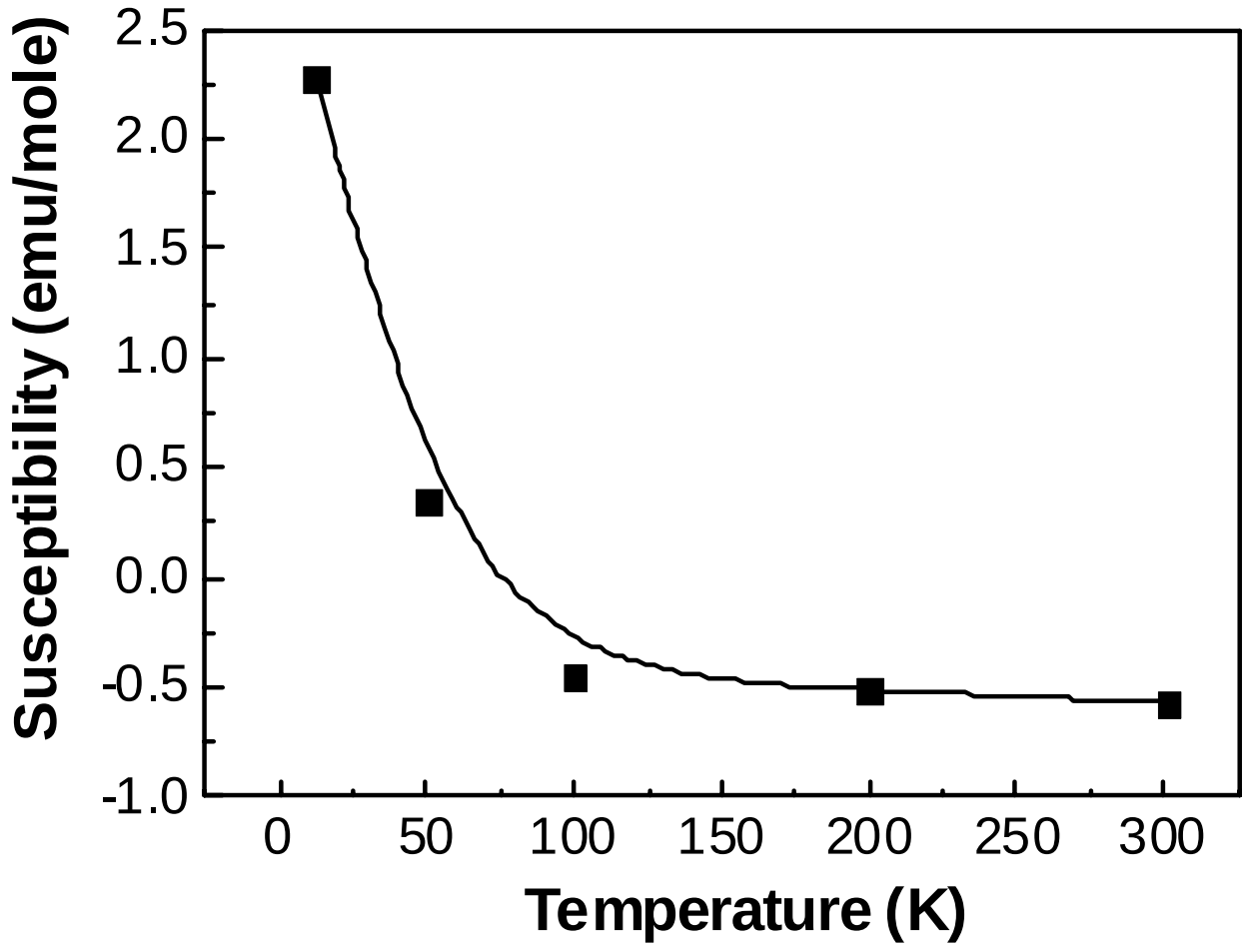

Fig. 3. Magnetic susceptibility versus temp of BFO nanorod.

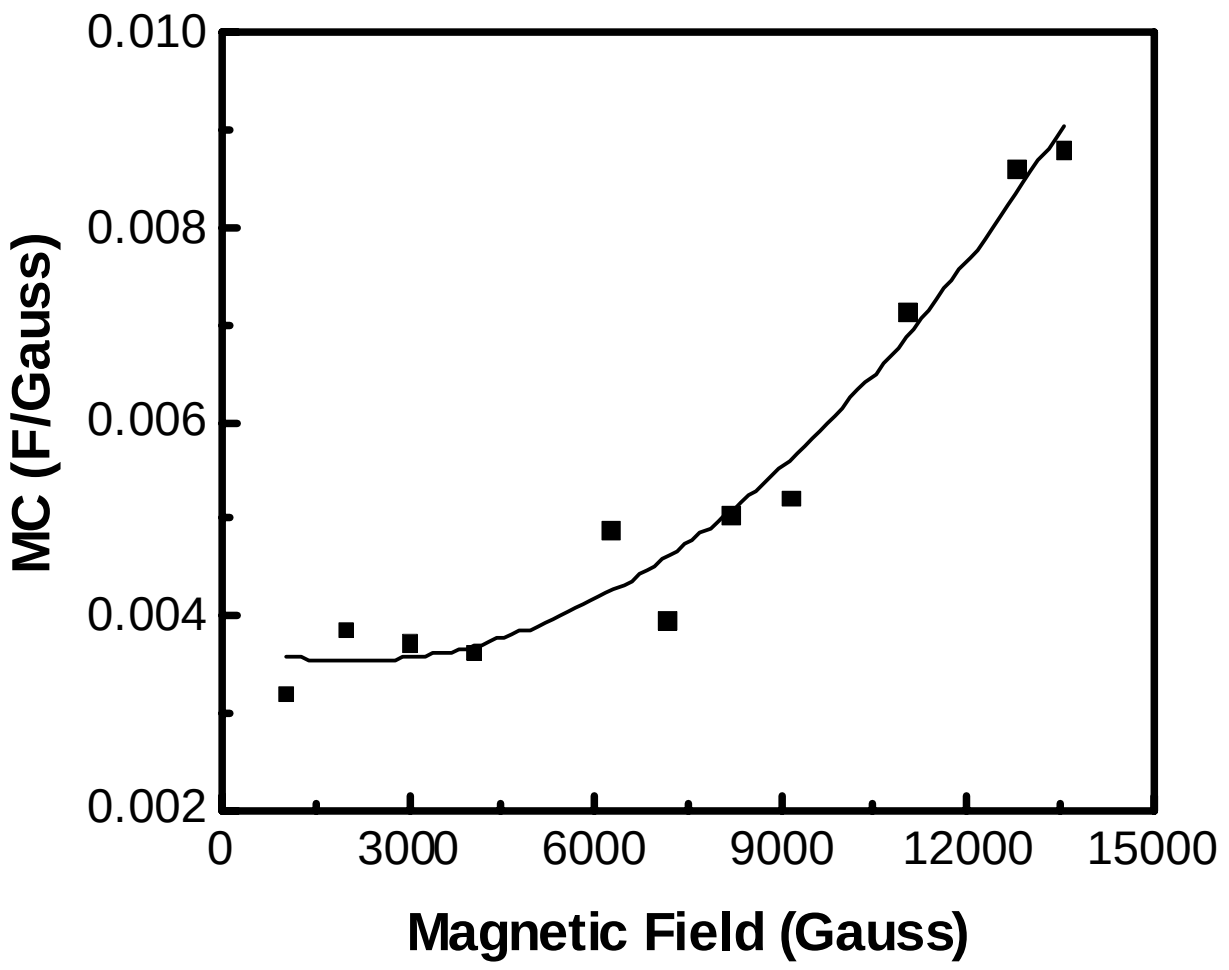

Fig. 4. Magnetocapacitance plot of BFO nanorod.